\documentclass[%
 reprint,
 showpacs,
 showkeys,
 amsmath,amssymb,
 aps,
]{revtex4-1}
\usepackage{amsmath}
\usepackage{cases}
\usepackage{amssymb}
\usepackage{CJK}
\usepackage{graphicx}
\usepackage{dcolumn}
\usepackage{bm}
\usepackage{color}
\usepackage[pagewise]{lineno}

\begin{document}
\begin{CJK*}{GBK}{} 

\preprint{APS/123-QED}

\title{Isotopic ratio, isotonic ratio, isobaric ratio and Shannon information uncertainty}
\thanks{Supported by National Natural Science Foundation of China under Grant
No. 10905017, Program for Science \& Technology Innovation Talents in Universities
of Henan Province (13HASTIT046), and the Young Teacher Project in Henan Normal
University.} %

\author{Chun-Wang MA
}\email{Corresponding author: machunwang@126.com\\}
\author{Hui-Ling WEI
}
\affiliation{%
Institute of Particle and Nuclear Physics, Henan Normal University, Xinxiang, 453007 China 
}%



\date{\today}

\begin{abstract}
The isoscaling and the isobaric yield ratio difference (IBD) probes, which both are constructed
by yield ratio of fragment, provide cancelation of parameters. The information entropy theory is
introduced to explain the physical meaning of the isoscaling and IBD probes. The similarity between
the isoscaling and IBD results is found, i.e., the information uncertainty determined by the IBD
method equals to $\beta-\alpha$ determined by the isoscaling [$\alpha$ ($\beta$) is the parameter
fitted from the isotopic (isotonic) yield ratio].
\end{abstract}
\maketitle
\end{CJK*}

\section{introduction}
In heavy-ion collisions (HICs) above intermediate energy, nuclear matters from
sub-saturation to supra-saturation densities can be produced, which makes HICs
be a unique experimental method to study the abnormal nuclear matters on earth.
The supra-saturation nuclear matter, which is produced in the compression of the
overlapping zone of the projectile and target nuclei, can not be probed directly.
The supra-saturation nuclear matters produced in HICs are related to dense nuclear
matters in compact astronomical body like neutron star, which attracts much interest
both theoretically and experimentally. On the theoretical side of studying the HICs,
the descriptions of the compressing and expanding of the collisions, at the same
time the decay of the hot fragments still face many challenges. On the experimental
side, the whole processes of the reaction like a black-box, with only the emitted
light particles and final residues measurable. Most of the probes detecting the
processes of HICs are based on the measurable light particles or final fragments \cite{BALi08PR}.
But the final fragments carry only part information of the initial collisions since
they undergo the decay process. Depending on the density and temperature, the nuclear
symmetry energy is one of the important properties of nuclear matter Since nuclear
symmetry energy can not be measured directly, the many results of nuclear symmetry
energy, which are extracted based on different indirect probes, are in conflict. Till
now, the nuclear symmetry energy is still an open question in nuclear physics, and it
is still important to find new probes to study the nuclear symmetry energy \cite{BALi08PR,ChLWFront07,NatowitzPRL10}.

The isoscaling method, which uses the isotopic or isotonic yield ratio, is one of the
important methods to study the nuclear symmetry energy of the sub-saturation nuclear
matter produced in HICs \cite{Huang10Mscaling,HShanPRL,Isoscaling}. The isobaric ratio
methods, which use the isobaric yield ratios, have been proposed to study the nuclear
symmetry energy of finite nuclei \cite{Huang10IYR,MaIYR11PRC06,Huang-PRC11,NST13Lin,MaIYR12CPL06,MaIYR12EPJA06,MaIYR13CPC,NST13WADA,PMar12PRCIsob-sym-isos,MaIYR12NPR,RCIMa14},
the chemical potential difference between neutrons and protons \cite{IBD13PRC,IBD13JPG},
and the density difference between projectiles \cite{IBD14Ca}. The volume effects manifested
in the results \cite{Souza12finite} are found to originated from the neutron-skin of
neutron-rich fragments\cite{IBFinite13-1,IBFinite13-2}. Besides, the ratios of fragments
are also used to detect the temperature of the reaction \cite{AlbNCA85DRT,MaCW13CTP,MaCW12PRCT,MaCW13PRCT,Ma2013NST}.
A systematic comparison between the results of the isoscaling and the isobaric yield ratio
difference (IBD) methods proves that the results of isoscaling and IBD are similar \cite{IBD13PRC,IBD13JPG}.
In both the isoscaling and IBD methods, the yield ratios of fragments provide cancelations
of special terms or parameters influencing the yield of fragment, which facilitates the
study of nuclear symmetry energy \cite{Huang10IYR,MaIYR11PRC06,MaIYR12EPJA06}.

The Shannon information theory is a method to measure the uncertainty in a random variable
which quantifies the expected value of the information contained in a message, and can extract
reliable information in the information transition from measured observable \cite{Shannon,Jaynes,IFEtrpApp,Fran}.
The Shannon information theory has many similarity compared to the black-box characteristics
of the HICs processes. The ideas of Shannon information entropy has been introduced to study
the hadron decaying branching process \cite{EntrHC}, and probe the liquid-gas transition in
the disassemble of the colliding system in HICs \cite{YGMaZipfPRL99}. In this article, we will
introduce the information entropy theory to understand the isoscaling and IBD probes.

\section{Shannon Information entropy Theory}

In the Shannon information theory, considering a system which has multi events $S=\{e_1,e_2, \cdots, e_n\}$
with the corresponding probability $\{p_1, p_2, \cdots, p_n\}$, the information uncertainty of a
certain event $e_i$ (or the information $e_i$ contained) is
defined as,
\begin{equation}\label{ShInfoUn}
U(e_i)=-\mbox{ln}p_i,
\end{equation}
with $U(e_i)$ in units of $nats$. 

If the probability of the event is non-uniform, the information entropy
of the system can be defined as \cite{Shannon,Jaynes,EntrHC,YGMaZipfPRL99},
\begin{equation}\label{Entp-sys}
H(S) =-\sum_{i=1}^{n}p_i\mbox{ln}p_i.
\end{equation}
The information entropy and information uncertainty can be used interchangeably \cite{Jaynes}.
In some applications, $U(e_i)$ is also named as information entropy of one event. To
differ the concept of the previously defined information entropy in Refs. \cite{EntrHC,YGMaZipfPRL99},
$U(e_i)$ is called as the \textit{information uncertainty}. In HICs, all the types of
particles and fragments form a system. Each type of particle and fragment can be viewed
as an independent event with different probability denoted by yield or cross section
($\sigma$). In the work of Y. G. Ma, the liquid-gas transition is found in the
information entropy of the system \cite{YGMaZipfPRL99}. In this work, we concentrate
on the information uncertainty of the final fragments, which is believed to carry
part of the information of reactions.

\section{Isotopic/Isotonic/Isobaric Ratio and Information Uncertainty}

In the free energy based theories describing the HICs above the Fermi energy, the yield
of a fragment follow the exponential function, which is mainly decided by the free energy, the
chemical properties of the source, and temperature \cite{Tsang07BET,Huang10IYR,Huang-PRC11,MFM1,GrandCan,Karth12BE_T}.
In the ratio of fragment yield, the cancelation of some terms makes it possible to study
the retained terms in the exponential function, and specific physical parameters can be
studied. Some probes based on yield ratios are proposed, for examples, the isotopic temperature
probes \cite{AlbNCA85DRT,MaCW13CTP} and the isobaric temperature probes \cite{MaCW12PRCT,MaCW13PRCT,Ma2013NST},
the isoscaling probe for nuclear symmetry energy \cite{HShanPRL,Huang10Mscaling}, and the
isobaric yield ratio difference probes for nuclear symmetry energy \cite{Huang10IYR,MaIYR12EPJA06,IBD13PRC,IBD13JPG,IBD14Ca},
etc. We will explain the physical meaning of the result for the isoscaling and isobaric
ratios using the information uncertainty theory.

Assuming the thermal equilibrium, in the grand-canonical ensembles theory within the
grand-canonical limit, the yield of a fragment is given by \cite{GrandCan,Tsang07BET},
\begin{equation}\label{yieldGC}
\sigma(A, I) = CA^{\tau}exp\{[F(A,I)+\mu_{n}N+\mu_{p}Z]/T\},
\end{equation}
where $A$ and $I = (N - Z)$ denote the mass and neutron-excess of the fragment; $C$
depends on the reaction system; $F(A, I)$ is free energy of the fragment, and $\mu_n$
($\mu_p$) denotes the chemical potential of the neutrons (protons). In the modified
Fisher model, the yield of a fragment is described in a similar form by considering
the entropy of exchanging the neutrons and protons \cite{MFM1,Huang10IYR}. The
exponential law of the fragment yield makes it easy to explain the probes based on
the fragment yield by using the information uncertainty theory since in Eq. (\ref{ShInfoUn})
the logarithm operation of yield probability frees the parameters in the
yields in Eq. (\ref{yieldGC}).

\subsection{Isotopic \& Isotonic Ratios}

For the isoscaling method, which uses the isotopic ratio and isotonic ratios, we denote a
fragment with neutron numbers $N$ and proton numbers $Z$ as $(N, Z)$ for convenience. From
Eq. (\ref{yieldGC}), which represents the residue probability for a fragment, the information
uncertainty included in the fragment can be written as,
\begin{equation}\label{InfoFrag}
U(e)=-\mbox{ln}\sigma(N, Z),
\end{equation}
with $e$ denoting an event corresponding to the fragment $(N, Z)$. In reactions of similar
measurements, the temperatures of the specific fragment can be assumed as the same, thus
the free energies of $(N, Z)$ equal in the two reactions. The difference between the information
uncertainty of fragment $(N, Z)$ in the two reactions can be written as,
\begin{equation}\label{Infoisoscaling}
\Delta_{21} U(e)=U_2(e)-U_1(e) =\mbox{ln}\sigma_{1}(N, Z)-\mbox{ln}\sigma_{2}(N, Z),
\end{equation}
Inserting Eq. (\ref{yieldGC}) into (\ref{Infoisoscaling}), one obtains,
\begin{eqnarray}\label{DInfoScaling}
&\Delta_{21} U(e) \hspace{6.8cm}\nonumber\\
&=c1-c2+[F_{1}(N, Z)-F_2(N, Z)+N(\mu_{n1}-\mu_{n2})\nonumber\\
&+Z(\mu_{p1}-\mu_{p2})]/T   \nonumber\\
                &=\Delta c+[\Delta F_{12}(N,Z)+N\Delta \mu_{n12}+Z\Delta \mu_{p12}]/T,
\end{eqnarray}
$\Delta c=c1-c2$, with 1 and 2 are indexes denoting the reactions and the corresponding parameters.
Defining $\alpha=\Delta \mu_{n21}/T =(\mu_{n2}-\mu_{n1})/T$ and $\beta=\Delta \mu_{p21}/T=(\mu_{p2}-\mu_{p1})/T$,
Eq. (\ref{DInfoScaling}) can be written as,
\begin{equation}\label{DfragUe}
\Delta_{21} U(e)=\Delta c - N\alpha - Z\beta
\end{equation}
with $\alpha$ ($\beta$) being the fitting parameter from the isotopic (isotonic) ratio between
reactions. For isotopic ratio, $Z\beta$ is a constant (labeled as $C_z$). From Eq. (\ref{DfragUe}),
the following can be obtained,
\begin{equation}\label{DInfoIsotope}
\Delta_{21} U(e_p)=\Delta c + C_z + N\Delta \mu_{n12}/T.
\end{equation}
Similarly, for isotonic ratios,  $N\alpha$ is a constant (labeled as $C_n$).
From Eq. (\ref{DfragUe}), the following can be obtained,
\begin{equation}\label{DInfoIsotone}
\Delta_{21} U(e_n)=\Delta c + C_n + Z\Delta \mu_{p12}/T,
\end{equation}
with $e_p$ and $e_n$ denoting the isotopic and isotonic events. It is shown that $\Delta_{21} U(e_n)$
depends on the reaction systems due to $\Delta c_{12}$. But only $\alpha$ and $\beta$ are the interested
parameters. For isotopic (isotonic) ratio, the $C_z$ ($C_n$) is assumed to be a constant. This assumption
can only be fulfilled when the nuclear density does not change.

One fragment belongs both to an isotopic chain and an isotonic chain. In the isoscaling
analysis, the fragment is related to $\alpha$ in its isotopic ratio and to $\beta$ in its
isotonic ratio simultaneously. The difference between the information uncertainty included
in a fragment from its isotopic ratio and isotonic ratio is,
\begin{equation}\label{DifInfoanbp}
\Delta_{12} U(e_p)-\Delta_{12} U(e_n)= N\alpha - Z\beta +C_z - C_n= 0.
\end{equation}
If $C_z = C_n$ can be fulfilled, one has,
\begin{equation}\label{DifInfoaapprxb}
\frac{\alpha}{\beta}=\frac{Z}{N}.
\end{equation}
This can only happen in the neutron-proton symmetric matter.

\subsection{Isobaric Ratio}
For isobaric ratio, the fragment will be denoted as $(A, I)$. The information uncertainty
difference between the isobars differing 2 in $I$ can be written as,
\begin{equation}\label{InfoIBorg}
\Delta U(e_b)_{(I+2, I)}=\mbox{ln}\sigma(A, I)-\mbox{ln}\sigma(A, I+2),
\end{equation}
where $e_b$ denotes the isobaric event. Inserting Eq. (\ref{yieldGC}) into Eq. (\ref{InfoIBorg}), the
$CA^{\tau}$ term cancels out and one obtains,
\begin{equation}\label{DInfoIB}
\Delta U(e_b)_{(I+2, I)}=[\Delta F(A, I, I+2)-\mu_{n}+\mu_{p}]/T,
\end{equation}
with $\Delta F(A, I, I+2)=F(A, I)-F(A, I+2)$. Assuming the temperatures of two reactions are the
same, one can define the difference between the information uncertainty of isobars,
\begin{equation}\label{DInfoIB}
\Delta_{21} U(e_b)_{(I+2, I)}=[\Delta_{21} F(A, I, I+2)-\Delta\mu_{n21}+\Delta\mu_{p21}]/T,
\end{equation}
$\Delta_{21} F(A,I,I+2)=0$ can be assumed, which results in the following equation,
\begin{equation}\label{DInfoIBIS}
\Delta_{21} U(e_b)_{(I+2, I)}=(-\Delta\mu_{n21}+\Delta\mu_{p21})/T=\beta-\alpha,
\end{equation}
$\Delta\mu_{n21}$ ($\Delta\mu_{n21}$) is the same as in Eq. (\ref{DInfoScaling}), which
denotes the difference between the chemical potential of neutrons (protons) of the two reactions.
In Eq. (\ref{DInfoIBIS}), the correlation between the isobaric ratio difference and the
isoscaling parameters $\alpha$ and $\beta$ is explicated. This correlation has also been
illustrated and verified experimentally \cite{IBD13PRC,IBD13JPG}.

\subsection{Discussion}

Both the isoscaling and isobaric methods use the yields of fragments produced in two similar
reactions. In the isoscaling and IBD methods, the
free energies of the fragments cancel out in different manners by assuming the same temperatures
of the reactions. In the isotopic (isotonic) ratios, the constant $\Delta_{12} c$ in Eqs.
(\ref{DInfoIsotope}) and (\ref{DInfoIsotone}) makes the difference between the information
uncertainty of the isotopic (isotonic) ratios depends on the reaction system, but it is unimportant
in the isoscaling analysis since it only cares $\alpha$ and $\beta$. In the isobaric ratio, the
cancelation of $CA^{\tau}$ makes it convenient to compare the fragment yield in reactions besides
those induced by isotopic projectiles or on isotopic targets \cite{IBD13PRC,IBD13JPG,MaIYR12EPJA06}.
In the real reactions, the yield of fragment sometimes does not obey the isoscaling, in which
case the isoscaling analysis encounters difficulties. The isobaric ratio, which uses only two
or three isobars, does not require regular distributions of fragments as in the isoscaling method.

When comparing the information uncertainty included in the isoscaling and IBD probes, it should
also be pointed out that the IBD results are obtained directly from the fragment ratio, while
the isoscaling results are obtained indirectly since $\alpha$ and $\beta$ are the fitting parameters
from the isotopic or isotonic ratio. From the information theory, the IBD probe has advantages
compared with isoscaling, and it should be more sensitive to the change of the reactions.

\section{Summary}

The information entropy theory is introduced to explain the isoscaling and IBD probes. The
physical meanings of the isoscaling and IBD probes, which both use fragment ratio to make
cancelation of parameters, are explained in the information uncertainty manner. The similarity
between the isoscaling and IBD results is found, i.e., the information uncertainty determined
by the IBD method equals to the value of $\beta-\alpha$. The IBD probe is shown to have advantage
to the isoscaling method both in experiment and theoretical analysis, which could also be used
when the fragment does not obey the isoscaling.

\subsection*{Acknowledgment}
We thank Prof. Ma Yugang, at Shanghai Institute of Applied Physics, Chinese Academy of Sciences),
for the useful discussion of the information entropy theory.


\begin{thebibliography}{}
%
%
\bibitem{BALi08PR}
B.-A. Li, L.-W. Chen, C. M. Ko %
Phys. Rep. \textbf{464} (2008) 113.
\bibitem{ChLWFront07} L. W. Chen \textit{et al.}, Front. Phys. China \textbf{2} (2007) 327.

\bibitem{NatowitzPRL10}
J. B. Natowitz, G. R\"{o}pke, S. Typel 
{\it et al.},
Phys. Rev. Lett. \textbf{104} (2010) 252501.

\bibitem{Huang10Mscaling}
M. Huang, Z. Chen, S. Kowalski 
{\it et al.},
Nucl. Phys. \textbf{A847} (2011) 233. 
\bibitem{HShanPRL} H. S. Xu \textit{et al.}, Phys. Rev. Lett. \textbf{85} (2000) 716.

\bibitem{Isoscaling} Y. G. Ma \textit{et al.}, Phys. Rev. \textbf{C 69}, 064610 (2004); Phys. Rev. \textbf{C 72}, 064603 (2005).

\bibitem{Huang-PRC11}
M. Huang, A. Bonasera, Z. Chen 
{\it et al.},
Phys. Rev. \textbf{C81} (2011) 044618.

\bibitem{PMar12PRCIsob-sym-isos} 
P. Marini, A. Bonasera, A. McIntosh 
{\it et al.},
Phys. Rev. C \textbf{85} (2012) 034617.
\bibitem{NST13Lin}
W. Lin, R. Wada, M. Huang,  Liu X, Zhao M and Chen Z,
Nucl. Sci. Tech. \textbf{24} (2013) 050511.
\bibitem{NST13WADA}
R. Wada, M. Huang, W. Lin, Liu X, Zhao M, Chen Z,
Nucl. Sci. Tech. \textbf{24} (2013) 050501.

\bibitem{Huang10IYR}
M. Huang, Z. Chen, S. Kowalski 
{\it et al.},
Phys. Rev. {\bf C81} (2010) 044620.

\bibitem{MaIYR11PRC06}
C. W. Ma, F. Wang, Y. G. Ma, and C. Jin,
Phys. Rev. {\bf C83} (2011) 064620.

\bibitem{MaIYR12EPJA06}
C. W. Ma, J. Pu, H. L. Wei, S. S. Wang, H. L. Song, S. Zhang, and L. Chen,
Eur. Phys. J {\bf A48} (2012) 78.

\bibitem{MaIYR12CPL06}
C. W. Ma, J. Pu, S. S. Wang, and H. L. Wei,
Chin. Phys. Lett. {\bf 29} (2012) 062101.

\bibitem{MaIYR13CPC}
C. W. Ma, H. L. Song, J. Pu 
{\it et al.},
Chin. Phys. \textbf{C37} (2013) 024102.

\bibitem{RCIMa14}
C.-W. Ma, S.-S. Wang, Y.-L. Zhang, H.-L. Wei, arXiv:1402.5493 [nucl-th].
\bibitem{MaIYR12NPR}
J. Pu \textit{et al.},
Nucl. Phys. Rev. \textbf{29} (2012) 129 (in Chinese).

\bibitem{IBD13PRC}
C. W. Ma, S. S. Wang, Y. L. Zhang, H. L. Wei, Phys. Rev. \textbf{C87} (2013) 034618.
\bibitem{IBD13JPG}
C. W. Ma, S. S. Wang, Y. L. Zhang, H. L. Wei,
J. Phys. G: Nucl. Part. Phys. \textbf{40} (2013) 125106.

\bibitem{IBD14Ca}
C. W. Ma, J. Yu, X. M. Bai, Y. L. Zhang, H. L. Wei, S. S. Wang,
Phys. Rev. \textbf{C 89} (2014) 057602.


\bibitem{Souza12finite} 
S. R. Souza and M. B. Tsang,
Phys. Rev. C \textbf{85} (2012) 024603.

\bibitem{IBFinite13-1}
C. W. Ma, S. S. Wang, H. L. Wei, and Y. G. Ma,
Chin. Phys. Lett. \textbf{30} (2013) 052101.
\bibitem{IBFinite13-2}
C. W. Ma, H. L. Wei, Y. G. Ma,
Phys. Rev. \textbf{C88} (2013) 044612.


\bibitem{AlbNCA85DRT}
S. Albergo 
{\it et al.},
Nuovo Cimento {\bf A89} (1985) 1.

\bibitem{MaCW13CTP} 
C. W. Ma, \textit{et al.},
Commun. Theo. Phys. \textbf{59} (2013) 95.

\bibitem{MaCW12PRCT} 
C. W. Ma, J. Pu, Y. G. Ma, R. Wada, S. S. Wang,
Phys. Rev. \textbf{C86} (2012) 054611.


\bibitem{MaCW13PRCT}
C. W. Ma, X. L. Zhao, J. Pu 
{\it et al.},
Phys. Rev. \textbf{C88} (2013) 014609.

\bibitem{Ma2013NST}  
C. W. Ma, C. Y. Qiao, S. S. Wang, F. M. Lu, L. Chen, and M. T. Guo,
Nucl. Sci. Tech. \textbf{24} (2013) 050510.


\bibitem{Shannon}
C. E. Shannon,
Bell System Technical Journal {\bf 27(3)}, 379 (1948).

\bibitem{Fran}
G. Francois and O. Stefano, (2008) Entropy methods for the Boltzmann equation: lectures
from a special semester at the Centre \'{e}mile Borel, Institut H. Poincar\'{e},
Paris, 2001. Springer. p. 14. ISBN 978-3-540-73704-9. (doi:10.1007/978-3-540-73705-6)

\bibitem{Jaynes}
E. T. Jaynes, Phys. Rev. {\bf 106(4)}, 620 (1957).

\bibitem{IFEtrpApp}
K. G. Denbigh and J. S. Denbigh, \textit{Entropy in Relation to Uncomplete Knowledge} Cambridge, 1985. Cambridge University Press. ISBN: 978-0521256773.
\bibitem{EntrHC}
P. Brogueira, J. Dias de Deus, and I. P. da Silva, Phys. Rev. \textbf{D 53}, 5283 (1996);
Z. Cao and Rudolph C. Hwa, Phys. Rev. \textbf{D 53}, 6608 (1996).
\bibitem{YGMaZipfPRL99}
Y. G. Ma, Phys. Rev. Lett. \textbf{83}, 3617 (1999).

\bibitem{MFM1} 
R. W. Minich, S. Agarwal, A. Bujak 
{\it et al.},
Phys. Lett. \textbf{B 118} (1982) 458.
\bibitem{Karth12BE_T}  
C. Karthikraj, N. S. Rajeswari, and M. Balasubramaniam, Phys. Rev. C \textbf{86} (2012) 014613.

\bibitem{Tsang07BET}
M. B. Tsang, W. G. Lynch, W. A. Friedman 
{\it et al.},
Phys. Rev. \textbf{C76} (2007) 041302(R).
\bibitem{GrandCan} 
C. B. Das, S. Das Gupta, X. D. Liu, and M. B. Tsang
Phys. Rev. \textbf{C64} (2001) 044608.

\end{thebibliography}
\end{document}